\documentclass[journal=jpca,manuscript=article
]{achemso}

\usepackage{graphicx} 
\graphicspath{{Figures/}}
\usepackage[caption=false]{subfig}
\usepackage{float} 
\usepackage{dcolumn}  
\usepackage{bm}       

\usepackage{array}
\usepackage{rotating}
\usepackage{outlines}
\usepackage{latexsym}
\usepackage{amsfonts}
\usepackage{amssymb}
\usepackage[usenames,dvipsnames,table]{xcolor}
\usepackage{tikz}

\newcommand{\etal}{\emph{et al.}{} }

\definecolor{ultrapink}{rgb}{1.0, 0.44, 1.0}
\definecolor{purplepizzazz}{rgb}{1.0, 0.31, 0.85}
\newcommand{\red}[1]{\textcolor{black}{#1}}

\usepackage[version=4]{mhchem} 

\DeclareUnicodeCharacter{0303}{\textasciitilde}
\DeclareUnicodeCharacter{03A3}{$\Sigma$}
\DeclareUnicodeCharacter{03A0}{$\Pi$}

\author{O. A. Krohn}
\affiliation{Department of Physics, University of Colorado, Boulder, Colorado, USA\red{, 80309}}
\alsoaffiliation{JILA, National Institute of Standards and Technology and the University of Colorado, Boulder, Colorado, USA\red{, 80309}}
\author{K. J. Catani}
\affiliation{Department of Physics, University of Colorado, Boulder, Colorado, USA\red{, 80309}}
\alsoaffiliation{JILA, National Institute of Standards and Technology and the University of Colorado, Boulder, Colorado, USA\red{, 80309}}
\alsoaffiliation{Laboratory for Atmospheric and Space Physics, University of Colorado, Boulder, CO, USA\red{, 80303}}
\author{S. P. Sundar}
\affiliation{Department of Chemical Engineering, The University of Melbourne, Parkville 3010, Victoria, Australia}
\author{J. Greenberg}
\affiliation{Department of Physics, University of Colorado, Boulder, Colorado, USA\red{, 80309}}
\alsoaffiliation{JILA, National Institute of Standards and Technology and the University of Colorado, Boulder, Colorado, USA\red{, 80309}}
\author{G. da Silva}
\affiliation{Department of Chemical Engineering, The University of Melbourne, Parkville 3010, Victoria, Australia}
\author{H. J. Lewandowski}
\affiliation{Department of Physics, University of Colorado, Boulder, Colorado, USA\red{, 80309}}
\alsoaffiliation{JILA, National Institute of Standards and Technology and the University of Colorado, Boulder, Colorado, USA\red{, 80309}}
\email{lewandoh@colorado.edu}

\begin{tocentry}
\includegraphics[width=8.0cm]{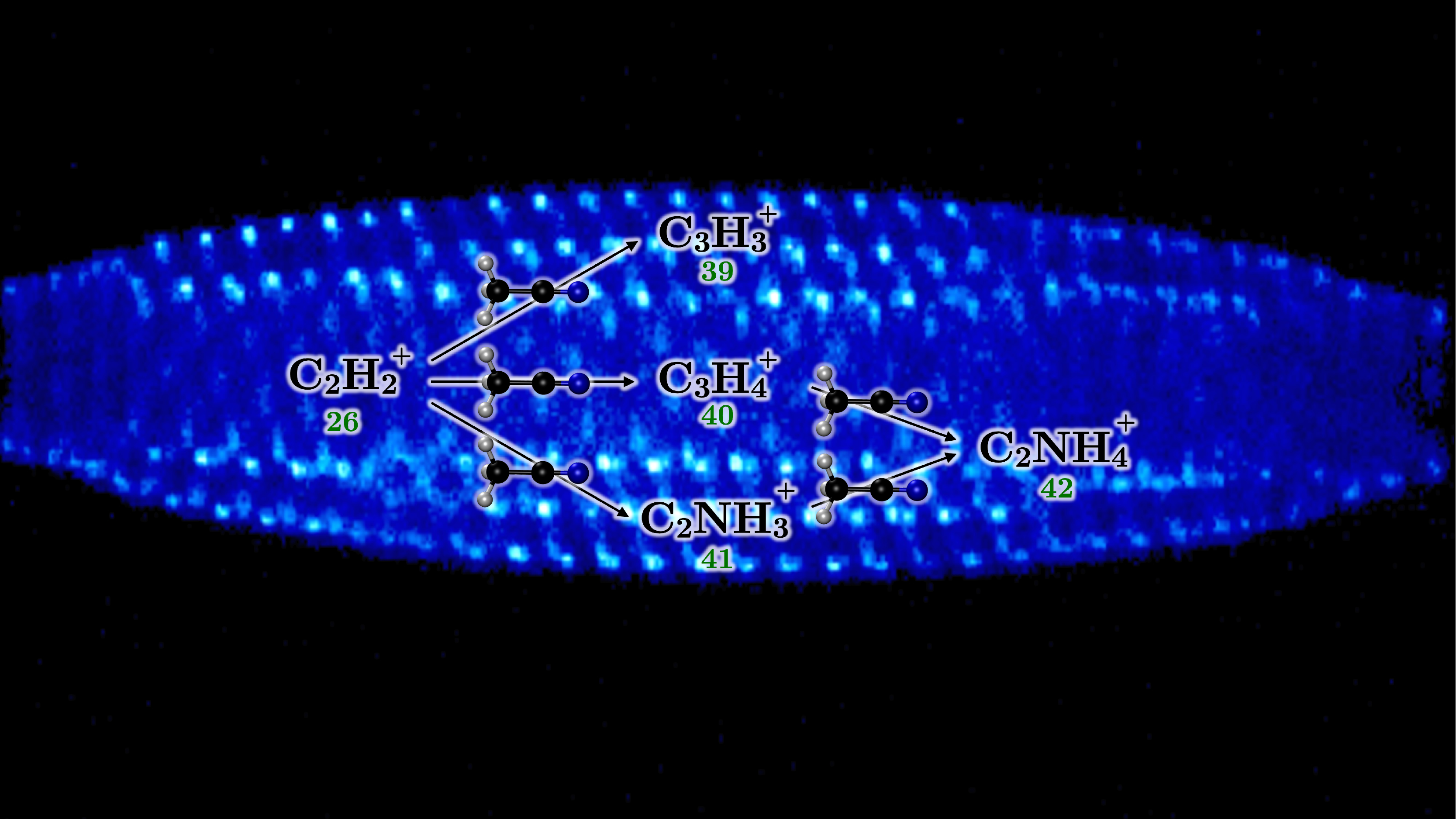}
\end{tocentry}

\title[]
  {Reactions of Acetonitrile with Trapped, Translationally Cold Acetylene Cations}

\begin{document}

\date{\today}

\begin{abstract} 
The reaction of the acetylene cation (\ce{C2H2+}) with acetonitrile (\ce{CH3CN}) is measured in a linear Paul ion trap coupled to a time-of-flight mass spectrometer. \ce{C2H2+} and \ce{CH3CN} are both noted for their astrochemical abundance and predicted relevance for understanding prebiotic chemistry. The observed primary products are \emph{c}-\ce{C3H3+}, \ce{C3H4+} and \ce{C2NH3+}. The latter two products react with excess \ce{CH3CN} to form the secondary product \ce{C2NH4+}, protonated acetonitrile. The molecular formula of these ionic products can be verified with the aid of isotope substitution via deuteration of the reactants. Primary product reaction pathways and thermodynamics are investigated with quantum chemical calculations and demonstrate exothermic pathways to two isomers of \ce{C2NH3+}, two isomers of \ce{C3H4+}, and the cyclopropenyl cation \emph{c}-\ce{C3H3+}. This study deepens our understanding of the dynamics and products of a pertinent ion-molecule reaction between two astrochemically abundant molecules in conditions that mimic those of the interstellar medium.
\end{abstract}

\maketitle

\section{Introduction}
Small carbonaceous species, in both their neutral and ionic forms, are essential building blocks in several diverse chemical environments. Acetylene (\ce{C2H2}) and its cation \ce{C2H2+} are ubiquitous in fuels, flames, planetary environments, and in the interstellar medium (ISM).\cite{Waite2007,Goodings1982,Frenklach:2002,Bohme1992,Herbst2009ism} Much of the early understanding of \ce{C2H2+} reactivity was developed from mass spectrometry studies of acetylene-rich flames, motivated by the hypothesis that ion-neutral reactions contribute to soot formation.\cite{Goodings1982,Keil:1985,Calcote:1990,Calcote:1988,Calcote:1981,Fialkov:2013,Hayhurst:1987kc,Semo:1984uv} Far from these hot and dense  environments, interest in \ce{C2H2+} has more recently been focused on its role in colder and less dense regimes like planetary atmospheres and the ISM. Small cations of this type are thought to have a role in ion-neutral condensation reactions that may lead to the formation of larger more complex organic species.\cite{Bohme1992,Herbst2009ism,Schiff1979} Understanding the reactivity of \ce{C2H2+} in a controlled laboratory setting under low temperature and pressure conditions is crucial to understanding its reactivity in diverse chemical environments such as the ISM and planetary atmospheres. 

Nitriles, including acetonitrile (\ce{CH3CN}), are pervasive in many regions of space and have been tied to several areas of complex chemistry taking place in the ISM. Particularly interesting are the ion-neutral reactions that form dense haze layers of Titan, which may have implications for prebiotic chemistry.\cite{Waite2007,ANICICH2000,Cravens:2014} \ce{CH3CN} itself has been identified in the ISM,\cite{Solomon1971} with a notable presence in cold dark clouds,\cite{Matthews1983} low mass protostars,\cite{Codella2009, Bisschop2008} and hot cores.\cite{Cazaux_2003, bisschop2007,Gerin1992} Furthermore, it has been detected in the dust from several comets, including Halley,\cite{Kissel1987} Hale-Bopp, (C/1995 O1)\cite{Biver1997} and, more recently, 67P/Churyumov-Gerasimenko.\cite{Morse2019} Further understanding the reactivity of this prevalent neutral with a fundamental carbocation, like \ce{C2H2+}, is important for many areas of chemistry. 

Reactions of \ce{C2H2+} have been previously studied using several different techniques and neutral reactants.\cite{Greenberg2021,schmidgreenberg2020,Myher1968,IRAQI1990,PALM200631,Jarrold1983,Anicich2006} Recent work from our group showed the different mechanisms of \ce{C2H2+} reacting with two structural isomers of \ce{C3H4}, and demonstrated how isotopic substitution is a powerful tool for determining chemical reaction processes.\cite{Greenberg2021,schmidgreenberg2020} The specific reaction of \ce{C2H2+ + CH3CN} (subject of the current study) was previously measured at room temperature using a selected ion flow tube (SIFT) apparatus.\cite{IRAQI1990}
The reported reaction products were \ce{C2NH4+}, \ce{C3H5+}, \ce{C3H4+}, and the \ce{C4NH5+} adduct (putatively [\ce{C2H2}$\cdot$\ce{CH3CN}]$^+$) with nearly equal branching. This study was conducted at high pressure, which can stabilize highly excited reaction complexes through collisions with background buffer gas. Ion-neutral chemical reactions typically produce such unstable complexes, which are unlikely to stabilize in regions like the ISM. The low pressure and collision energy regime of the current study should more closely mimic conditions present in the ISM and yield a better understanding of the reactivity and dynamics of these two species in these remote domains.

In particular, ion traps and Coulomb crystals have been fruitful environments to study a myriad of gas-phase chemical reactions and interesting quantum phenomena.\cite{Toscano2020,heazlewoodChapter, Puri2019,bell2009,Okada2015, molhave2000} In these experiments, atomic ions (here \ce{Ca+}) are trapped and directly laser-cooled, forming Coulomb crystal structures that sympathetically cool co-trapped ions to translational temperatures below 10\,K. This type of experimental setup allows for controlled collisions of purified ionic species with neutral molecules, and is particularly suited for long interrogation times. Furthermore, coupling a linear Paul ion trap (LIT) to a time-of-flight mass spectrometer (TOF-MS) allows for the exact determination of the molecular weight and number of the chemical species present in the trap at high resolution. This high resolution is more than enough to detect differences of a single mass unit for the range of masses studied here.\cite{SchmidRSI2017} The conditions created in these types of experiments are relevant to the cold and low density conditions of space. Additionally, the low collisional energies ($\sim$\,100\,K) impose stricter bounds on reaction energies, which yield valuable comparisons with quantum chemical computational modeling. 

Here, we use a LIT TOF-MS apparatus to characterize the reactions of \ce{C2H2+ + CH3CN} under low pressure and temperature conditions. The primary products are found to be \emph{c}-\ce{C3H3+}, \ce{C3H4+}, and \ce{C2NH3+}, which are unambiguously assigned using isotope substitutions and quantum chemical calculations. This study provides insight into the reactivity of two astrochemically abundant molecules, \ce{C2H2+} and \ce{CH3CN}, and additional formation pathways of fundamental carbocations \ce{H3C3H+}, \ce{CH2CCH2+} and \emph{c}-\ce{C3H3+}, which could prove useful for refining chemical models that rely on accurate energetics and electronic structure information. This work also provides useful information on the dissociation of excited [\ce{C4H5N+}]* cation, which is relevant to the relatively well-studied decomposition of the pyrrole cation [global minimum on the potential energy surface (PES)] in the photoionization of neutral pyrrole.\cite{RENNIE2010142,doi:10.1021/ja00542a018,RENNIE1999217} Observing and studying the interaction of these two important interstellar species in a cold and low pressure regime is consequential to understanding ion-neutral chemistry in extraterrestrial environments.

\section{Methods}

\subsection{Experimental Methods}
Kinetic data are measured using a LIT radially coupled to a TOF-MS. The LIT TOF-MS has been described in detail elsewhere\cite{SchmidRSI2017,krohn2021,schmidgreenberg2020,Greenberg2021} and only a brief summary of the features pertinent to the current experiment are given here. Acetylene cations are produced using a (1+1) resonance-enhanced multiphoton ionization scheme. A $\sim$2\% mixture of \ce{C2H2} or \ce{C2D2} (CDN isotopes 99\%-d2) seeded in He is expanded supersonically to create a molecular beam, which passes through a skimmer into the center of the trap, where it is overlapped with a focused beam from the output of a frequency-doubled pulsed dye laser (216\,nm for \ce{C2H2} or 218\,nm for \ce{C2D2};\cite{vancraen1985a,orr19951+} LIOPTEC LiopStar; 10\,ns pulse, 1\,mJ/pulse). Small amounts of contaminant ions are formed in the ionization process. These unwanted ions are ejected from the trap by sweeping over resonance frequencies of the specific mass-to-charge ratio (\emph{m/z}) of undesired ions.\cite{Roth2007,Schmidt2020}

\ce{Ca+} ions are subsequently loaded into the trap with the acetylene ions by non-resonantly photoionizing calcium from a resistively heated oven, using the third harmonic of an Nd:YAG (355\,nm; Minilite, 10\,Hz, $\sim7$\,mJ/pulse). The resulting \ce{Ca+} ions are laser-cooled using two external cavity diode lasers (397 and 866\,nm). The cold \ce{Ca+} sympathetically cool the co-trapped acetylene ions via Coulomb interactions, forming a mixed Coulomb crystal structure. The reaction experiments are visually monitored by collecting \ce{Ca+} ion fluorescence with a microscope objective, which focuses the light onto an intensified CCD camera located above the trap. The lighter acetylene ions, which do not fluoresce, arrange themselves in the center of the trap as a cylindrical dark core within the \ce{Ca+} ions. A typical experiment utilizes 150-300 acetylene ions trapped with $\sim1000$ \ce{Ca+} ions, all of which are translationally cold ($\sim10$\,K), where the temperature of the ions is limited by micromotion heating. \red{This loading process takes about a minute, which allows the acetylene ions sufficient time to relax from any possible vibrational excitation that may have occurred in the REMPI process.}

Once acetylene and \ce{Ca+} ions are loaded into the trap, neutral acetonitrile [9-10\% \ce{CH3CN} or \ce{CD3CN} (Cambridge Isotopes 99.8\%-d3) in \ce{N2}] is leaked into the vacuum chamber ($3\times10^{-9}$\,Torr or $4\times10^{-7}$\,Pa gas pressure at 300\,K) for a set duration of time using a pulsed leak-valve (LV) scheme.\cite{Jiao1996,SchmidMOL2019} The typical chamber base pressure is $6\times10^{-10}$\,Torr ($8\times10^{-8}$\,Pa) and the measurements of gas pressures in the chamber are recorded using a Bayard-Alpert hot cathode ionization gauge. The opening of the LV defines the zero-time point; the LV remains open for several different time steps between 0-400\,s before the ions are ejected into the TOF-MS. \red{The TOF-MS has a resolution of m/$\Delta$m $\geq$ 1100, which can resolve neighboring masses with excellent accuracy and precision.\cite{SchmidRSI2017}} This process is repeated about 12 times for every time step and measured ion numbers from each mass channel are averaged over each time step\red{, including the zero-time point, which measures the initial number of ions in the trap.} The average number of reactant and product ions are then normalized by the initial acetylene ion numbers and plotted against time, giving a reaction curve. These reaction curves are then used to determine the relevant rates of the reactions. Reaction curves are collected in the same manner for all isotopologues, such that reactions with all four possible combinations of isotopologues are measured. Due to the excellent mass resolution of the TOF-MS,\cite{SchmidRSI2017} we are able to observe mass shifts from these small substitutions. TOF-MS traces are tracked for all of the ionic species present and the total number of ions are compared at each time point to ensure that the numbers are constant throughout the experiment. This rules out systematic losses of ions from the trap. Figures illustrating conservation of charge for each reaction are given in the Supplementary Information (SI). 

\subsection{Computational Methods}
Quantum chemical calculations are carried out using the Gaussian 16 program.\cite{g16} Geometry optimizations and harmonic vibrational frequency calculations are performed at the M06-2X/6-31G(2df,p) level of theory. Single point energies are computed using the G3X-K method\cite{DASILVA2013109}, which is specifically developed for thermochemical kinetics and is accurate to within 0.03\,eV on average for barrier height predictions. Harmonic vibrational frequency calculations of optimized minima and transition states affirm the presence of zero and one imaginary frequency, respectively. The inter-connectivity of transition states is in all cases confirmed by intrinsic reaction coordinate (IRC) calculations.

\section{Results \& Discussion}
\subsection{Reaction measurements}\label{ExpRes}
\begin{figure}[h]
    \centering
    \includegraphics[width=8.5cm]{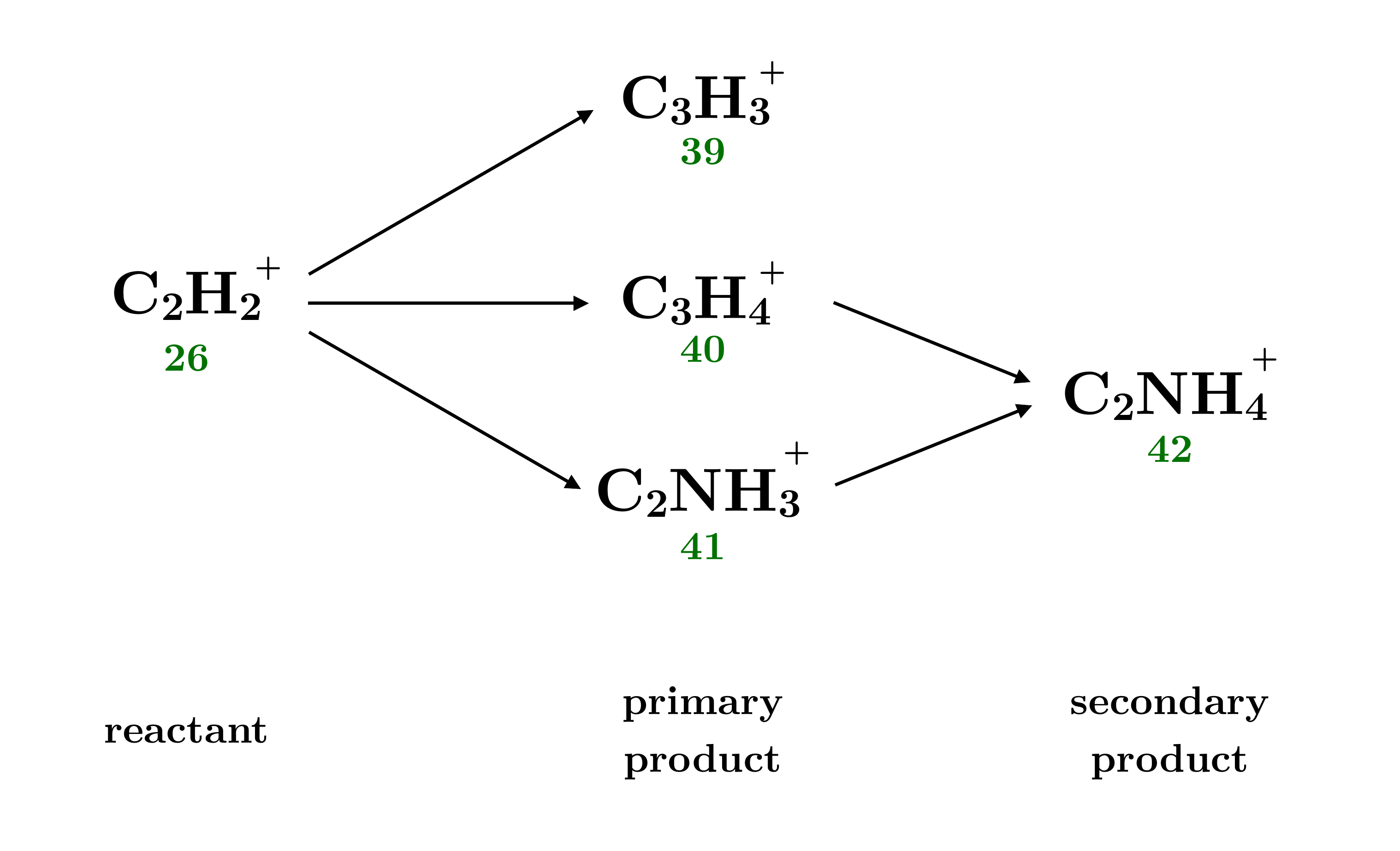}
    \caption{Model for reaction of \ce{C2H2+ + CH3CN}, \emph{m/z} ratio (blue number) and determined chemical formula below the molecule. The reaction order of each molecule is located at the bottom of the figure. Each arrow represents a reaction with a neutral \ce{CH3CN} molecule.} 
    \label{fig:model}
\end{figure}

The discussion here is limited to the specific reaction of \ce{C2H2+ + CH3CN}. The results of the other three isotopologue reactions are discussed in terms of product identification. Details of these isotopologue reactions are reported in the SI.
\begin{figure}[h]
    \centering
    \includegraphics[width=8.5cm]{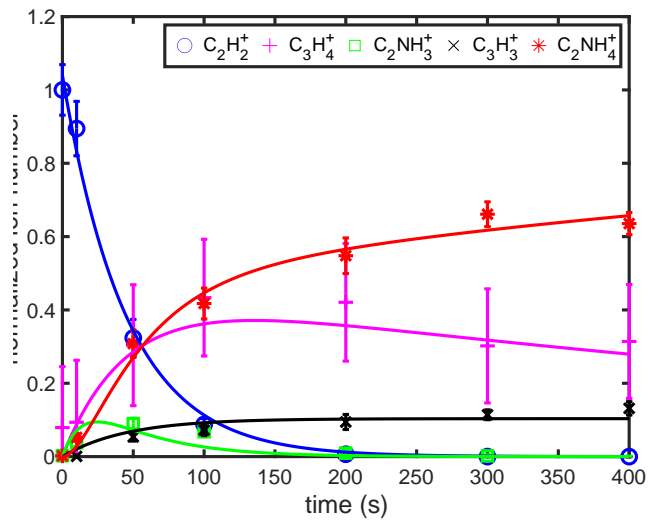}
    \caption{Measured ion numbers of \ce{C2H2+} (blue\,\textcolor{blue}{$\bm{\circ}$}), \ce{C3H4+} (magenta\,\textcolor{purplepizzazz}{$\bm{+}$}), \ce{C2NH3+} (green\,\textcolor{green}{$\bm{\Box}$}), \emph{c}-\ce{C3H3+} (black\,{$\bm{\times}$}), and \ce{C2NH4+} (red\,\textcolor{red}{$\bm{\ast}$}) as a function of time. Data are normalized by the initial ion number of \ce{C2H2+} ($\sim 200$). Each data point represents the mean and standard error from twelve experimental runs per time point. The averaged data are fit using a pseudo-first-order reaction rate model (solid lines).} 
    \label{fig:reactioncurve}
\end{figure} 
The reaction of \ce{C2H2+ + CH3CN} produces the curves shown in Fig. \ref{fig:reactioncurve}. Here, \ce{C2H2+} (blue $\bm{\circ}$, \emph{m/z} 26) reacts away over time to produce the primary products \emph{c}-\ce{C3H3+} (black $\bm{\times}$, \emph{m/z} 39), \ce{C3H4+} (magenta $\bm{+}$, \emph{m/z} 40) and \ce{C2NH3+} (green $\bm{\Box}$, \emph{m/z} 41). The reduction in the ion numbers of \ce{C2H2+} coincides with the growth of the three primary products. These primary products reduce over time as the population of secondary product \ce{CH3CNH+} (red $\bm{\ast}$, \emph{m/z} 46) increases from reactions with excess neutral \ce{CH3CN}. \ce{C2NH4+} is confirmed as a secondary product as its numbers continue to increase after all the \ce{C2H2+} has completely reacted. \ce{C2NH4+} is a product from reactions of \ce{C3H4+ + CH3CN} and \ce{C2NH3+ + CH3CN}. These observed products are used to construct a kinetic model (see Fig. \ref{fig:model}) in order to fit the reaction data, as well as extract reaction rate coefficients and product branching ratios. The experimental conditions are such that \ce{CH3CN} is in excess throughout the course of the reaction, which is represented by a pseudo-first-order kinetic model. This model uses a set of differential equations (given in the SI) to fit the experimentally observed ion numbers as a function of time. The resulting fits are shown as lines in Fig. \ref{fig:reactioncurve}. The decay rate of \ce{C2H2+} is extracted from this fit and can be used to calculate the reaction rate constant, \emph{k} = 4.5$\pm$0.6 $\times 10^{-9}$\,cm$^{3}$/s. This value is obtained by measuring the partial pressure of acetonitrile gas with a hot cathode ion gauge close to the trapping region as the reaction proceeds. Hot cathode ion gauges of this type are subject to systematic uncertainties at pressures below $1\times10^{-8}$\,Torr \red{ associated with the nonlinear sensitivity of ion gauges in this regime}.\cite{jousten:2007} Reaction rate constants for the isotopologue combinations are similar to that of the fully hydrogenated reaction, with exact values reported Tab. S1 in the SI.

Reaction curves are also measured for the four unique pairs of isotopologue reactants (see SI for these reaction curves). These kinetic data are used to confirm the product assignments from the \ce{C2H2+ + CH3CN} reaction. Primary product mass distributions for each isotopologue combination and branching ratios are given in Tab. \ref{tab:branching}. These branching ratios were determined by dividing the growth rate of each respective \red{$m/z$ channel} by the total acetylene ion loss rate. 

\red{While an analysis of the branching ratios across the four different reactions would be interesting, the available isotope options for each product  coincide with each other and hinder such analysis. For example, in the \ce{C2D2+ +CH3CN} reaction case, the \ce{C3H_xD_y+}($x+y=3)$ reaction product can be found in $m/z$ channel 39, 40 or 41. The \ce{C3H_xD_y+}($x+y=4$) product in this case can feasibly be $m/z$ 41 or 42. Additionally, \ce{C2NH_xD_y+}($x+y=3$) can be $m/z$ 41, 42, or 43. A very similar case emerges for \ce{C2H2+ +CD3CN}. Even without the expected splitting in the fully deuterated case, two of the three primary products coincide in the same channel ($m/z$ 44). This overlap in possible product masses precludes any direct comparison of branching for a single \textit{product} across the different reaction sets. Indeed, trying to compare a single \textit{mass channel} for the reaction sets is misleading. For example, in the fully hydrogenated case $m/z$ 41 is solely \ce{C2NH3+}. This channel could be \ce{C3HD2+} or \ce{C3H3D+} for the \ce{C2D2+ +CH3CN} reaction and must be \ce{C3HD2+} for \ce{C2H2+ +CD3CN}. None of our assigned products can be $m/z$ 41 when they are fully deuterated (not seen). The fact that the branching differs in the $m/z$ 41 channel for the four reactions is the natural result of the statistical $m/z$ options available to each product due to hydrogen-deuterium swapping in the reaction complex.} 

\begin{table*}
    \centering
    \caption{Branching ratios for the primary products of acetylene cations reacting with acetonitrile. The numbers are given in percentages and uncertainties are derived from the 90\% confidence interval from the pseudo-first-order model fits.}
    \label{tab:branching}
    \begin{tabular}{lcccccc}
    \hline\hline
       Reactants  & \emph{m/z} 39 &  \emph{m/z} 40 & \emph{m/z} 41 & \emph{m/z} 42 &\emph{m/z} 43 & \emph{m/z} 44  \\
       \hline
     \ce{C2H2^+ + CH3CN} & 10(2)& 43(10) & 47(6)&  \\ 
     \ce{C2D2^+ + CH3CN} & 3(3) & 38(18) & 22(15) & 14(13) & 23(14) &  \\
        \ce{C2H2^+ + CD3CN} & & 26(12) & 4(3) & 17(11) & 53(21) &  \\
        \ce{C2D2^+ + CD3CN} && 55(14) &  & 10(4) &  & 34(8) \\
     \hline\hline
    \end{tabular}
\end{table*}

\red{While assessment of the isotopologue effects on branching ratios is not possible for this reaction, these branching ratios are a resource for identifying our products.} In the fully deuterated case, \ce{C2D2+ + CD3CN}, primary products shift, $m/z~40 \rightarrow~44$ (\ce{C3D4+}), $m/z~41 \rightarrow~44$ (\ce{C2ND3+}), and $m/z~39 \rightarrow~42$ (\ce{C3D3+}). The secondary product also shifts in the fully deuterated reaction, $m/z~42 \rightarrow~46$ (\ce{C2ND4+}). \red{This analysis can be checked for each reaction set.} In each case, mass shifts are consistent with the products observed in the titular reaction. \red{A specific example of this point involves our reassignment of the $m/z$ 41 channel to \ce{C2NH3+} in the fully hydrogenated case. This channel could theoretically be \ce{C3H5+}. This would require the product masses to shift entirely to the $m/z$ 44 channel in the \ce{C2H2+ +CD3CN} case, which is not observed. In addition, this would require a $m/z$ 46 primary product in the fully deuterated case, where we only see the second-order product, \ce{CD3CND+}.} Product assignments are further supported by quantum chemical calculations, in particular calculated reaction thermodynamics discussed below.

The primary product \ce{C3H4+} is partially obscured by overlap with mass coincident and more abundant \ce{Ca+} signal. Because of this overlap, it is important to understand how calcium is reacting with \ce{CH3CN} in order to correctly model the formation and depletion of \ce{C3H4+}. The reaction of \ce{Ca+ + CH3CN} has been previously studied,\cite{okadaPRA2013} and only produces \emph{m/z} 66  \ce{CaCN+}.\cite{krohn2021} We verified that there were no interfering mass products with the current reaction by reacting \ce{Ca+} with \ce{CH3CN} without any acetylene present, and concluded no statistically significant differences of the formation of \ce{CaCN+} to the reaction when acetylene cations are present. In addition to reactions with \ce{CH3CN}, some minute amounts of gaseous \ce{H2O} are present in our system that react with \ce{Ca+} to form \emph{m/z} 57 \ce{CaOH+}. This reaction has also been characterized under similar conditions.\cite{Okada2002} Again, we verified no statistically significant difference between reactions of \ce{Ca+ + H2O} with or without acetylene in the trap. These two reactions, \ce{Ca+} with both \ce{CH3CN} and \ce{H2O} are included in the model for \emph{m/z} 40, and details of this model and the specific differential equations for the fits are given in the SI. Both of these \ce{Ca+} reaction products were taken into account when calculating the reaction rates and branching ratios for the \emph{m/z} 40 channel. 

An insight into the dissociation of excited \ce{C4H5N+} also comes from numerous prior studies on pyrrole cation dissociation, which is an intermediate in the \red{dissociative} photoionization of neutral pyrrole.\cite{RENNIE1999217,RENNIE2010142} These studies identify \ce{C3H4+} (m/z 40) + HCN and \ce{C2NH3+} (m/z 41) + \ce{C2H2} as the experimentally observed products with their appearance energies at -1.68 eV and -1.81 eV, respectively in the reference to the reactant's energy. As mentioned above, the reaction of \ce{C2H2+ + CH3CN} has been previously measured by Iraqi \etal at room temperature and much higher pressures using a SIFT apparatus.\cite{IRAQI1990} They reported primary products \ce{C2NH4+}, \ce{C3H5+}, \ce{C3H4+}, and the adduct [\ce{C2H2}$\cdot$\ce{CH3CN}]$^+$ with nearly equal branching. 
In the SIFT study, \ce{C2NH4+} was assigned as a primary product, but in the current study, its formation clearly has a late onset that corresponds to the decrease in ion number of primary products \emph{m/z} 41 and \emph{m/z} 40 and is therefore re-assigned here as a secondary product. \ce{C2NH4+} is not observed in pyrrole cation decomposition studies. Furthermore, \emph{m/z} 41 is reassigned from \ce{C3H5+} to \ce{C2NH3+}. This is supported by the absence of \emph{m/z} 44 (\ce{C3H2D3+}) in the reaction of \ce{C2H2+ + CD3CN} \red{as previously mentioned}. Additionally, there appears to be no corresponding first order product at \emph{m/z} 46 (\ce{C3D5+}) in the fully deuterated reaction. It could be that, although the formation of \ce{C3H5+} is slightly exothermic ($\sim$\,-0.06\,eV), it might be impeded by a small barrier that is surmountable at 300\,K\,(26 meV), but not in the current study with less available energy. 

The adduct is not seen in the current study because we are in a low pressure regime and there are no collisions with buffer gas to quench possible intermediates like the adduct. Interestingly, \emph{c}-\ce{C3H3+} was not reported as a product in the SIFT study but experimentally detected in the pyrrole cation studies.\cite{RENNIE1999217} The SIFT study does show \emph{m/z} 40 as a product, also assigned to \ce{C3H4+}, further confirming this assignment without the mass-coincident \ce{Ca+}.

We are confident that we observe the formation of \ce{C3H4+} + \ce{HCN} in the reaction of \ce{C2H2+ + CH3CN}, as seen in the previous SIFT and in the pyrrole cation dissociation studies.\cite{IRAQI1990,RENNIE1999217,RENNIE2010142} However, we observe branching into the \emph{m/z} 40 channel (see Tab. \ref{tab:branching}) in the other data sets where one or both of the reactants is deuterated. We do expect a small amount of \emph{m/z} 40 to be present in the mixed data sets based on the observation of the primary product \emph{c}-\ce{C3H3+} \emph{m/z} 39 in the fully hydrogenated dataset. No \emph{m/z} 40 is predicted in the fully deuterated data set, yet a significant amount (about 43\%) of \ce{C2D2+} is converted to \emph{m/z} 40. We have dedicated significant effort to understanding the origin and possible identity of the \emph{m/z} 40 signal, and believe it to be a contaminant in our reactions. We verified experimentally that both reactants need to be present to see growth in \emph{m/z} 40 and that it is not the result of a reaction with \ce{Ca+}. 

Similar contamination issues have been observed before in the reactions of acetylene cations with different isotopologues of \ce{C3H4}. The contamination was pinpointed as impurities in the deuterated gases.\cite{schmidgreenberg2020,Greenberg2021} The current sample preparation of both \ce{CH3CN} and \ce{CD3CN} included upwards of 10 freeze-pump-thaw cycles that should help to purify the liquid samples, but this technique may not have been successful in eliminating contaminates. We used various samples of different levels of purity from several companies, as well as different mixing gases (Ar, \ce{N2}, and He), without any change in the observed branching into \emph{m/z} 40 in the fully deuterated data set. We also ran our experiments with trapped \ce{C2H2+} and \ce{C2D2+} and reacted with 100\% of the mixing gas (Ar, \ce{N2}, and He) and saw no growth in the \emph{m/z} 40 mass channel, indicating there was no contamination from the gas delivery and mixing process. Because both reactants must be present to see this growth, we can rule out reactions between trapped acetylene cations and its neutral counterpart, which may be ambient in the system. Likewise, for the same reasons, we are also able to rule out charge exchange between acetylene cations and ambient neutral \ce{Ca}.

We considered the possibility that \emph{m/z} 40 could consist of a different product than \ce{C3H4+}. However, the only plausible chemical formula for a primary product would be \ce{H2C2N+}, and all structural isomers for this chemical formula were found to be significantly endothermic ($>1.5$\,eV) at the G3X-K level of theory. We tested the hypothesis that some \emph{m/z} 39 product could be \ce{HC2N+}, which would convert to \emph{m/z} 40 in the fully deuterated case. However, all viable isomers for this chemical formula were also greater than 1.5\,eV endothermic and ruled out. The secondary product \ce{CN2+} was also considered as a hypothetical constituent of \emph{m/z} 40, however, it was also significantly endothermic and eliminated as a possibility. It is possible that the acetylene ions could be electronically excited from the initial ionization laser pulse. However, the lowest-lying electronic states all require two additional photons, with the exception of the lowest-lying doublet $^{2}\Sigma_{g}^{+}$, which lies 5.67\,eV above the ground state of \ce{C2H2+}.\cite{PERIC1998} This excited state is extremely short-lived ($<$ ns);\cite{GILLEN2001, CHAMBAUD1995} the ions have several seconds between loading and the introduction of the neutral reactant. For these reasons, we believe all the acetylene cations are in the ground electronic state. These observations, as well as comparisons with the previous ion trap and SIFT studies,\cite{schmidgreenberg2020,Greenberg2021,IRAQI1990} led us to believe that a contaminant is competing with the \ce{C2H2+ + CH3CN} reaction. The previous SIFT experiment showed \emph{m/z} 40 as a product and it was assigned as \ce{C3H4+}. We remain confident in our assignment of primary products in the reaction of \ce{C2H2+ + CH3CN}, and have exhausted the available resources in our current experimental set up to further purify or identify the possible contaminant. \red{Unfortunately, this further complicates quantitative analysis of the branching ratios. While the $m/z$ 40 channel in the fully hydrogenated case shows evidence of re-reacting to form a second order product (unlike the $m/z$ signal in all of the other reactions) not all of it reacts in our reaction time. It is possible that the contamination may only be associated with deuterated gasses, but we cannot omit the possibility that this contaminate is also present in the fully hydrogenated case. Nevertheless, the comparison of the reaction across the four isotopologue combinations yields experimental product assignments further verified by quantum chemical investigations and reaction thermodynamics, as will be discussed next.}

\subsection{Reaction thermodynamics}

The model of the reaction of \ce{C2H2+ + CH3CN} is given in Fig. \ref{fig:model}, and shows that as the two reactants come together, three primary products form, \emph{c}-\ce{C3H3+}, \ce{C3H4+} and \ce{C2NH3+}. Two of these primary products then go on to react with excess \ce{CH3CN} to form \ce{C2NH4+}, protonated acetonitrile. Since the \ce{C2H2+} ions are sympathetically cooled to 10\,K, reactions with room temperature (300\,K) \ce{CH3CN} result in a calculated collision energy characterized by a temperature of about 116\,K (10\,meV). This calculated collision energy provides an upper thermodynamic limit to the reaction. Two exothermic isomers were found as a possibility for the \ce{C3H4+} product, \ce{CH2CCH2+}, the allene cation, and \ce{H3C3H+}, the propyne cation, with hydrogen cyanide (HCN) as the corresponding neutral. This is in accordance with the previous Iraqi \etal SIFT and pyrrole cation dissociation studies.\cite{RENNIE1999217,RENNIE2010142} The charge transfer product \ce{CH3CN+} is not energetically viable compared to the \ce{C2NH3+} product. However, both isomers of \ce{C2NH3+}, \ce{H2CNCH+} (Eqn. \ref{p3}) and \ce{H2C2NH+} (Eqn. \ref{p4}) are significantly exothermic. \ce{C3H5+ + CN} is also exothermic, but as discussed above, the lack of a \emph{m/z} 44 product when reacting \ce{C2H2+ + CD3CN} indicates that \ce{C3H5+} is not observed in the current study and therefore is not modeled in the PES. \emph{c}-\ce{C3H3+} is identified as the cyclic isomer (cyclopropenyl cation) and is determined to be a primary product as all energetic limits for it as secondary product are endothermic (Eqs. \ref{p5}, \ref{s5}). The observed products are all exothermic with respect to the reactants computed at the G3X-K level of theory (see Eqs. \ref{p1}-\ref{s5}) and well under the energetic limit of the collision energy.

Primary products:
\begin{equation}\label{p1}
\begin{split}
\ce{C2H2+ + CH3CN -> CH2CCH2+ + HCN} \\
\Delta E = -1.57\,\text{eV}
\end{split}
\end{equation}
\begin{equation}\label{p2}
\begin{split}
\ce{C2H2+ + CH3CN -> H3C3H+ + HCN} \\
\Delta E = -0.91\,\text{eV}
\end{split}
\end{equation}
\begin{equation}\label{p3}
\begin{split}
\ce{C2H2+ + CH3CN -> H2CNCH+ + C2H2}\\
\Delta E = -1.08\,\text{eV}
\end{split}
\end{equation}
\begin{equation}\label{p4}
\begin{split}
\ce{C2H2+ + CH3CN -> H2C2NH+ + C2H2}\\
\Delta E = -1.56\,\text{eV}
\end{split}
\end{equation}
\begin{equation}\label{p5}
\begin{split}
\ce{C2H2+ + CH3CN -> }\text{\emph{c}-}\ce{C3H3+ + CNH2} \\
\Delta E = -0.37\,\text{eV}
\end{split}
\end{equation}

Secondary products:
\begin{equation}\label{s1}
\begin{split}
\ce{H3C3H+ + CH3CN -> CH3CNH+ + H2C3H} \\
\Delta E = -0.96\,\text{eV}
\end{split}
\end{equation}
\begin{equation}\label{s2}
\begin{split}
\ce{H3C3H+ + CH3CN -> CH3CNH+ + }\text{\emph{c}-}\ce{C3H3} \\
\Delta E = 0.54\,\text{eV}
\end{split}
\end{equation}
\begin{equation}\label{s3}
\begin{split}
\ce{H2CNCH+ + CH3CN -> CH3CNH+ + H2C2N} \\
\Delta E = -0.71\,\text{eV}
\end{split}
\end{equation}
\begin{equation}\label{s4}
\begin{split}
\ce{H2C2NH+ + CH3CN -> CH3CNH+ + H2C2N} \\
\Delta E = -0.23\,\text{eV}
\end{split}
\end{equation}
\begin{equation}\label{s5}
\begin{split}
\ce{H3C3H+ + CH3CN -> }\text{\emph{c}-}\ce{C3H3+ + CH3CHN} \\
\Delta E = 0.22\,\text{eV}
\end{split}
\end{equation}

\subsection{Reaction potential energy surface}
\begin{sidewaysfigure*}
    \centering
    \includegraphics[width=22cm]{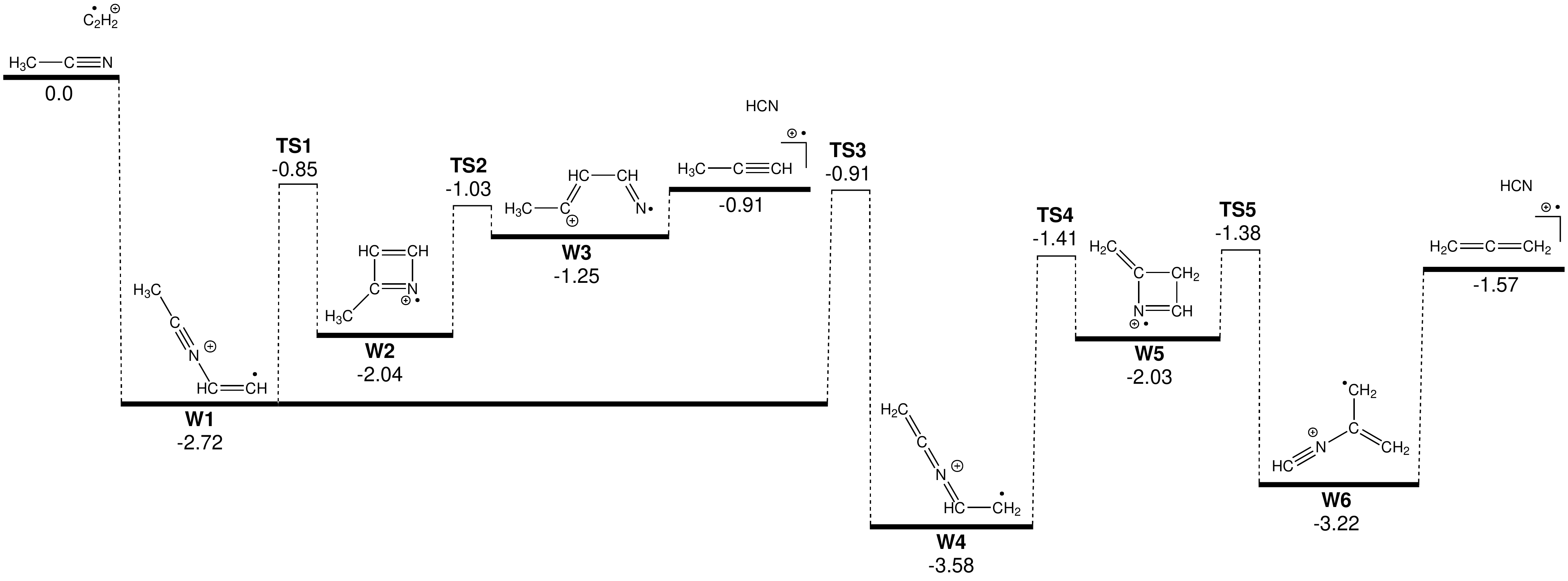}
    \caption{Potential energy surface (PES) for acetylene cation addition to acetonitrile to form \ce{CH2CCH+}/\ce{H3C3H+ + HCN}. Geometries were computed at the M06-2X/6-31G(2df,p) level, with energies calculated at the G3X-K level of theory. The energy values are 0 K enthalpies presented in eV.} 
    \label{fig:pes1}
\end{sidewaysfigure*}

\begin{figure*}[h]
    \centering
    \includegraphics[width=16cm]{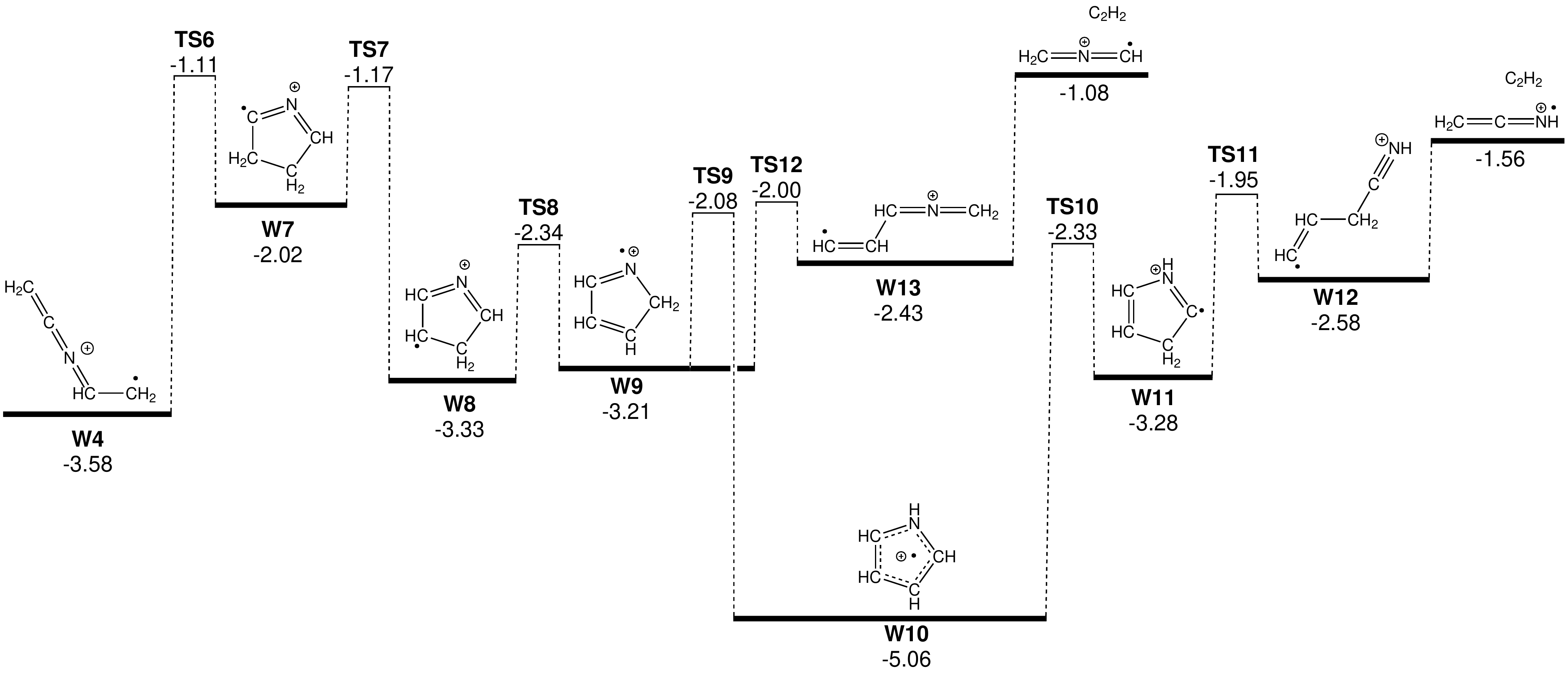}
    \caption{Potential energy surface (PES) diagram for the reaction channel forming \ce{H2CNCH+ + C2H2}. Geometries were computed at the M06-2X/6-31G(2df,p) level, with energies calculated at the G3X-K level of theory. The energy values are 0 K enthalpies presented in eV.} 
    \label{fig:pes2}
    \end{figure*}

\begin{figure}[h]
    \centering
    \includegraphics[width=8.5cm]{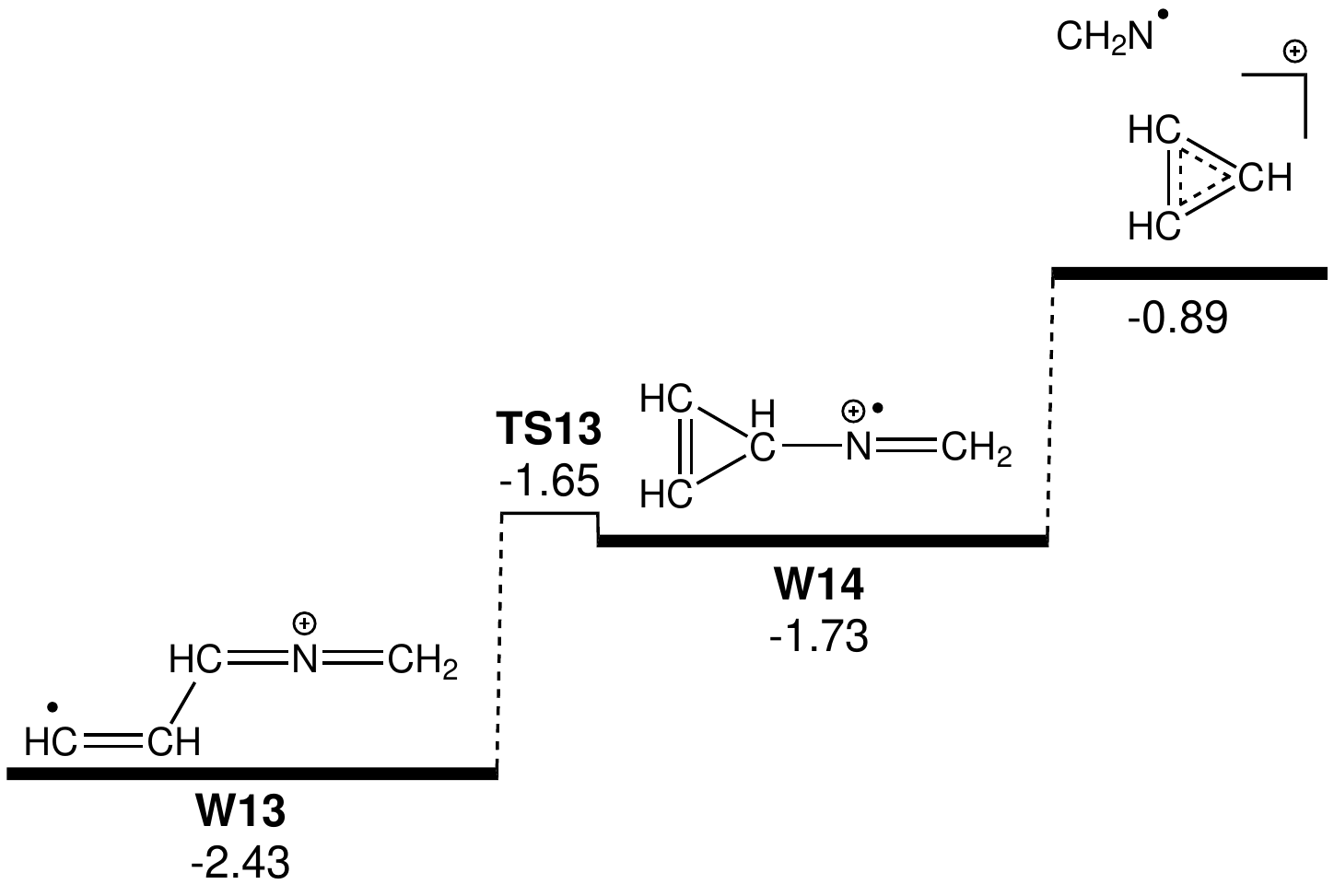}
    \caption{Potential energy surface (PES) diagram for the reaction channel forming \emph{c}-\ce{C3H3+ + CH2N}. Geometries were computed at the M06-2X/6-31G(2df,p) level, with energies calculated at the G3X-K level of theory. The energy values are 0 K enthalpies presented in eV.} 
    \label{fig:pes3}
    \end{figure}

Quantum chemistry calculations have been used to generate a PES, which connects the reactants to all observed products via multiple saddle points, show in Figs. \ref{fig:pes1}, \ref{fig:pes2} and \ref{fig:pes3}). In the developed \ce{C2H2+ + CH3CN} reaction mechanism, all transition states, intermediates, and product sets are exothermic compared to the reactants, such that the reaction complex can sample all of the stationary points before exiting the surface without the need for vibrational excitation above what is provided by ion-molecule complex formation. The PES not only identifies reaction pathways to possible product channels, but also provides a basis for understanding the kinetics of product formation. 

The section of the PES relevant to \ce{CH2CCH+}/\ce{H3C3H+} + HCN formation is presented in Fig. \ref{fig:pes1}. \ce{C2H2+} addition to \ce{CH3CN} commences with no entrance barrier to generate an adduct (W1), located 2.72\,eV below the reactants. Two competitive pathways exist from the well W1. 
The first channel gets to propyne cation by the formation of four membered ring intermediate (W2), by overcoming a 1.86\,eV energy barrier (0.85\,eV below the reactants). Subsequently, ring opening chemistry (TS2) takes place with a barrier of 1.01\,eV to form the intermediate W3. Again, all structures sit below the reactant energies [-1.03\,eV (TS2) and -1.25\,eV (W3)]. Finally, cleavage of a C-C bond in W3 generates the products \ce{H3C3H+ + HCN} at 0.91\,eV below the reactant energies.

The second competitive product channel from well W1 initiates via internal H–atom transfer (TS3) with a barrier of 1.81\,eV (0.91 below the reactants) to generate the intermediate W4. Multiple pathways exist from W4; the pathway to \ce{CH2CCH2+ (allene) + HCN} involves ring formation to W5 by crossing an energy barrier of 2.17\,eV (TS4). This is then followed by ring opening (TS5), forming W6. A C-N bond homolysis in W6 produces the dissociated products allene cation + HCN, represented in Fig. \ref{fig:pes1}.

The reaction pathway to the product channel \ce{H2CNCH+ or H2CCNH+ + C2H2} is represented in Fig. \ref{fig:pes2}. W4 undergoes C-C bond formation to produce a five-membered ring intermediate (W7) via a 2.46\,eV barrier (TS6). From W7, a series of H-shifts, produces W9 through the transition states TS7 and TS8 lying -1.17\,eV and -2.34\,eV below the reactant’s asymptote. The multiple isomer product channels that exist from W9 are represented in Fig. \ref{fig:pes2}. In the first channel, internal H-shifts occur in W9 through an energy barrier (TS9) of 1.13\,eV leading to the deepest well in the entire PES, W10 (pyrrole). The formed pyrrole undergoes an internal H-atom transfer (TS10) within the six-member ring by crossing an energy barrier of 2.73\,eV, producing W11. Finally, ring opening reaction in W11 leads to the product channel \ce{H2CCNH+ + C2H2} through the intermediate W12 via the transition state TS11 located -1.95\,eV below the reactant’s asymptote. 

The competitive channel from W9 consists of ring opening (TS12), crossing an energy barrier of 1.21\,eV, leading to W13. C–C bond homolysis in W13 leads to the product set \ce{H2CNCH+ + C2H2}. 

Another possible product channel from W13 is \emph{c}-\ce{C3H3+ + H2CN}. Whereby, W13 undergoes three-member ring formation leading to W14 through the transition state (TS13) lying -1.65\,eV below the reactant’s asymptote. C–N bond homolysis in W14 generates the product set \emph{c}-\ce{C3H3+ + H2CN}, represented in Fig. \ref{fig:pes3}.

Comparing the product channels identified here, all of the product sets are exothermic and proceed via transition states with all energies submerged below that of the reactants. For \ce{CH2CCH2+ + HCN} and \ce{H2CNCH+ + C2H2}, the products are  below the reactants (by 1.57 \,eV and 1.08\,eV) respectively, marginally more exothermic than the \ce{H3C3H+ + HCN}. The highest saddle point along this channel is TS3 (0.91\,eV below reactants); this is slightly above the energy of the exit channels. The rate constants and product branching fractions are decided based on the competition between energy and entropy factors. The entire reaction is complex and large parallel pathways exist from pyrrole cation. The dynamics will be the subject of future detailed kinetic studies. 
     
\section{Conclusion and outlook}
The gas-phase reaction of \ce{C2H2+ + CH3CN} is characterized in a low pressure and temperature regime using a LIT TOF-MS apparatus. Experimentally and computationally identified primary products are \emph{c}-\ce{C3H3+}, \ce{C2NH3+} and \ce{C3H4+}. Two products, \ce{C2NH3+} and \ce{C3H4+}, react with another acetonitrile to form the secondary product \ce{C2NH4+}. The experimental techniques used in this study provide low collision energies and low pressure environments, which limit the reaction dynamics to exothermic pathways and do not stabilize reactive intermediates. Additional reactions of \ce{C3D4+} with \ce{CH3CN} and \ce{CD3CN} would be useful to characterize, as it may provide further evidence that \ce{C3H4+} is indeed a primary product of the current reaction and may shed new light on \emph{m/z} 40 issues. This reaction has not been studied before in such a controlled manner and could be another important ion-neutral reaction in extraterrestrial environments. 

Future plans for the current setup of the LIT TOF-MS involve implementation of a 118\,nm light source, which would provide a cleaner ionization procedure for the creation of reactant ions.\cite{gray2021characterization} This vacuum ultra-violet light source would also enable detection of contaminants in reactant samples through photoionization experiments. We have also coupled the current apparatus to a traveling-wave Stark decelerator.\cite{osterwalder2010_twsd,perez2013_twsd,Shyur_2018b} This will extend the limits of cold, controlled reaction experiments by enabling control over the neutral reactant, both its quantum state and its velocity, leading to collision energies from 1-300\,K.\cite{greenberg2021decel} Having such reaction energy resolution and quantum control over both reactants will push the limits of our knowledge of fundamental chemical reactions and may provide new insights into astrochemical reactions.   


\section{Acknowledgments}
 This work was supported by the National Science Foundation (PHY-1734006, CHE-1900294) and the Air Force Office of Scientific Research  (FA9550-20-1-0323), as well as The University of Melbourne’s Research Computing Services and the Petascale Campus Initiative.



\providecommand{\latin}[1]{#1}
\makeatletter
\providecommand{\doi}
  {\begingroup\let\do\@makeother\dospecials
  \catcode`\{=1 \catcode`\}=2 \doi@aux}
\providecommand{\doi@aux}[1]{\endgroup\texttt{#1}}
\makeatother
\providecommand*\mcitethebibliography{\thebibliography}
\csname @ifundefined\endcsname{endmcitethebibliography}
  {\let\endmcitethebibliography\endthebibliography}{}

\end{document}